# Room-temperature continuous-wave Dirac-vortex topological lasers on silicon

Jingwen Ma[1,2,†], Taojie Zhou[2,3,†], Mingchu Tang[3,†], Haochuan Li[2], Zhan Zhang[2], Xiang Xi[1], Mickael Martin[4], Thierry Baron[4], Huiyun Liu[3], Zhaoyu Zhang[2,*], Siming Chen[3,*] and Xiankai Sun[1,*]

[1]*Department of Electronic Engineering, The Chinese University of Hong Kong, Shatin, New Territories, Hong Kong*

[2]*School of Science and Engineering, The Chinese University of Hong Kong, Shenzhen, Guangdong 518172, China*

[3]*Department of Electronic and Electrical Engineering, University College London, London, WC1E 7JE, United Kingdom*

[4]*Université Grenoble Alpes, CNRS, CEA-LETI, MINATEC, Grenoble INP, LTM, F-38054 Grenoble, France*

[†]*These authors contributed equally to this work*

[*]*Corresponding author: zhangzy@cuhk.edu.cn (Z.Z.); siming.chen@ucl.ac.uk (S.C.); xksun@cuhk.edu.hk (X.S.)*

**Robust laser sources are a fundamental building block for contemporary information technologies. Originating from condensed-matter physics, the concept of topology has recently entered the realm of optics, offering fundamentally new design principles for lasers with enhanced robustness. In analogy to the well-known Majorana fermions in topological superconductors, Dirac-vortex states have recently been investigated in passive photonic systems and are now considered as a promising candidate for single-mode large-area lasers. Here, we experimentally realize the first Dirac-vortex topological lasers in InAs/InGaAs quantum-dot materials monolithically grown on a silicon substrate. We observe room-temperature continuous-wave single-mode linearly polarized vertical laser emission at a telecom wavelength. Most importantly, we confirm that the wavelength of the Dirac-vortex laser is topologically robust against variations in the cavity size, and its free spectral range defies the universal inverse scaling law with the cavity size. These lasers will play an important role in CMOS-compatible photonic and optoelectronic systems on a chip.**

With the explosive growth of data traffic, it is highly desired to develop hybrid photonic integrated circuits (PICs) combining various optical components including lasers, modulators, waveguides, and detectors on a single chip[1]. Silicon is an outstanding material for PICs due to its unique strength in modulating, waveguiding, and detecting photons[2], but realizing high-power single-mode laser sources in silicon remains challenging[3]. Monolithic integration of III–V quantum-dot (QD) lasers on silicon[4, 5] is considered as a promising strategy to solve this problem because of its lower substrate cost, higher yield, and better CMOS compatibility compared with conventional heterogeneous integration methods[6]. Various III–V QD lasers have been demonstrated on silicon under room-temperature continuous-wave conditions, including distributed-feedback lasers[4, 7], ridge-waveguide lasers[5, 8], microring/microdisk lasers[9, 10], and photonic crystal cavity lasers[11]. In these lasers, the optical cavity determines the properties of the resonant modes and thus the lasing performance in many aspects. For instance, the free spectral range (FSR) of the cavity can determine laser emission in a single mode or in multiple modes. The size of the cavity determines the laser's effective modal volume ($V$), which further determines the maximal laser output power. Many practical applications require a high-power single-mode laser, pointing to the need for a cavity with a large FSR and $V$ simultaneously. However, this is challenging because most existing cavities suffer from a fundamental limitation that their FSR is inversely proportional to $V$.

Recently, topology as a mathematical concept has attracted intense interests in the realm of optics[12] and is revolutionizing the design strategies for lasers with many surprising properties[13-17]. Topological lasers have been demonstrated using zero-dimensional defect states in Su–Schrieffer–Heeger lattices[13, 18, 19] or corner states in high-order topological insulators[14, 20, 21]. Topological lasers have also been realized based on one-dimensional edge states in quantum Hall[15, 22] or quantum valley Hall[16, 23, 24] topological insulators. However, these topological lasers are not monolithically integrated on silicon substrates and cannot operate under room-temperature continuous-wave conditions, which strongly limit their potential applications in next-generation silicon-based PICs. Besides, these configurations either have limited scalability in modal volume $V$ (Ref. 13, 14, 18-20) or exhibit small FSR proportional to $V^{-1}$ (Ref. 15, 16, 22-24), which are undesired for practical lasers requiring single-mode operation over a large area. Fortunately, Dirac-vortex state[25], an analog of the well-known Majorana bound state (MBS) in superconductor electronic systems[26], has recently been discovered as a promising strategy to break this limitation. Such Dirac-vortex cavities possess a larger FSR (proportional to $V^{-1/2}$) than any other optical cavities known to date with a scalable modal volume, and thus are an ideal candidate for realizing high-power single-mode surface-emitting lasers. However, previous demonstrations of the Dirac-vortex cavities are limited to a passive photonic system[25]. Laser emission from an active Dirac-vortex cavity remains experimentally elusive.

Here, we experimentally demonstrated room-temperature continuous-wave Dirac-vortex topological lasers at a telecom wavelength using InAs/InGaAs QD materials monolithically grown on an on-axis silicon (001) substrate. Specifically, we designed and fabricated the Dirac-vortex photonic crystal lasers by harnessing an auxiliary orbital degree of freedom in topological insulators that we recently discovered in a nanomechanical system[27]. We observed single-mode linearly polarized vertical laser emission in such cavities under continuous-wave optical pump at room temperature. We compared the experimental far-field patterns with the simulated results and confirmed that the laser emission indeed occurs in the topologically protected MBS. Moreover, we fabricated the Dirac-vortex cavities with various cavity sizes and verified that their lasing wavelengths are always near the Dirac frequency due to the topological protection. Besides, we confirmed that the FSR of the Dirac-vortex lasers is unprecedentedly large and defies the universal inverse scaling law of $V^{-1}$. Our Dirac-vortex QD lasers not only are promising light sources for next-generation silicon-based PICs with topological robustness, but also open the door to exploration of various phenomena such as non-Hermiticity[28], bosonic nonlinearity[29], and quantum electrodynamics[30] in the context of topological MBS.

Figure 1a is a conceptual illustration of a fabricated Dirac-vortex topological laser based on InAs/InGaAs QDs epitaxially grown on an on-axis silicon (001) substrate. The photonic crystal was defined in the 362-nm-thick active layer and suspended by partially removing the 1-μm-thick $Al_{0.6}Ga_{0.4}As$ sacrificial layer. The active layer provides tight confinement to light in the vertical direction due to its high refractive index (~3.4). Figure 1b is a tilted-view scanning electron microscope image of the fabricated topological Dirac-vortex photonic crystal cavity. The active layer consists of two symmetric 40-nm-thick $Al_{0.4}Ga_{0.6}As$ cladding layers and four layers of InAs/$In_{0.15}Ga_{0.85}As$ dot-in-well structures separated by 50-nm GaAs spacer layers (see Extended Data Fig. 1). Figure 1c is a cross-sectional bright-field transmission electron microscope image of the four-stack InAs/InGaAs QD layers. To obtain these high-quality InAs/InGaAs QD layers, the growth process of the III–V buffer and defect-filter layers shown in Fig. 1a was carefully optimized to minimize the effects of lattice mismatch between the III–V materials and silicon substrate (see Methods).

Figure 2a is a scanning electron microscope image of the fabricated topological photonic crystal with a hexagonal lattice. Figure 2b shows the detailed structure in a unit cell. The lattice constant of the photonic crystal is $a_0 = 641$ nm. Each unit cell contains six triangular holes that can be classified into two groups due to the $C_3$ rotational symmetry. The relative distance from the centers of these triangular holes to the center of the unit cell is $d = a_0/\sqrt{3} - \delta_t$, where a nonzero $\delta_t$ breaks the $T_P$ translational symmetry of the crystal along the vector **P** (blue arrow in Fig. 2b). The sizes of the two groups of triangular holes are governed by parameters $(s_1, s_2) = (s_0 + \delta_i, s_0 - \delta_i)$, where $s_0 = 220$ nm is the average side length of the holes and $\delta_i$ breaks the $C_2$ inversion symmetry of the crystal. It is interesting to note that photonic crystals with nonzero values of $\delta_t$ and $\delta_i$ correspond respectively to quantum spin Hall[30] and quantum valley Hall[16] photonic topological insulators. The simulated eigenfrequencies of the bulk states at the $\Gamma$ point of the first Brillouin zone exhibit a doubly degenerate anisotropic cone-like dispersion relation in the parameter space defined by $\delta_t$ and $\delta_i$ (Fig. 2c). The polar coordinates $(\delta_0, \theta)$ can be defined according to $(\delta_t, \delta_i) = \delta_0(\alpha \cdot \sin\theta, \cos\theta)$ with $\alpha = 0.65$ (0.33) for $\delta_t > 0$ ($\delta_t < 0$), so that the opened bandgap at the $\Gamma$ point has less $\theta$ dependence (Extended Data Fig. 2). Here, $\theta$ represents an auxiliary orbital degree of freedom that can be used to construct the MBS. An effective bulk Hamiltonian mathematically identical to the Jackiw–Rossi model[26] was theoretically obtained:

$$H(\mathbf{k}) = v_D \cdot (\sigma_x k_x + \sigma_y k_y) + \frac{\Delta_0}{2}\sigma_z(\tau_x \cos\theta - \tau_y \sin\theta), \quad (1)$$

where $\sigma_x$, $\sigma_y$, $\tau_x$, and $\tau_y$ are the Pauli matrices, $\mathbf{k} = (k_x, k_y)$ is the wave vector, $v_D$ is the effective Fermi velocity near the $\Gamma$ point, and $\Delta_0 = \varsigma\delta_0$ is the opened bandgap and is proportional to geometric parameter $\delta_0$ with a proportionality constant $\varsigma$.

Figure 2d shows the detailed structure near the center of the fabricated photonic crystal, which is color-coded by the spatially varying parameters $\delta_0(R) = \delta_{max} \cdot [\tanh(R/R_0)]^4$ and $\theta(\varphi) = \varphi$. $R$ and $\varphi$ are the polar coordinates as marked in Fig. 2a, $R_0$ defines the size of the MBS, and $\delta_{max}$ controls the opened bulk bandgap based on $\Delta_{max} = \varsigma\delta_{max}$ at $R \gg R_0$. The modal profile ($z$ component of magnetic field) of the MBS was theoretically obtained:

$$h_z(R,\varphi) = g_0(R) \cdot |\psi_0\rangle, \quad (2)$$

which is determined by an envelope function $g_0(R)$ controlling the modal volume $V$ of the MBS, and a periodic Bloch mode $|\psi_0\rangle$ controlling the detailed modal profile in each unit cell (see Supplementary Information). The device in Fig. 2d adopted the parameters $R_0 = a$ and $\delta_{max} = 35$ nm. Through three-dimensional finite-difference time-domain simulation, we found that such a device supports a MBS in the bulk bandgap (Fig. 2e). The simulated modal profile of this topological MBS is mirror-symmetric with respect to the dashed gray line in Fig. 2f, because the device structure has the same type of symmetry. Note that the modal profiles of the MBS in our design are different from those based on the Kekulé distortion scheme due to the different choice of the Bloch mode $|\psi_0\rangle$. Compared with the Kekulé distortion scheme[25], ours is more suitable for practical lasers because it exhibits linearly polarized emission rather than vector-beam emission.

The microphotoluminescence (μ-PL) measurement was performed using a 632.8-nm continuous-wave pump laser at room temperature. The pump laser was focused onto the center of the Dirac-vortex cavity using a 50× object lens with a numerical aperture of 0.43. The light emitted from the topological cavity was collected by the same lens (see Methods). Figure 3a shows the measured spectra of a topological Dirac-vortex laser with structural parameters $R_0 = 2a$ and $\delta_{max} = 35$ nm. Increasing pump power $P_{pump}$ leads to enhanced light emission, with a peak wavelength $\lambda$ at ~1344 nm. The difference between the measured lasing wavelength ($\lambda = 1344$ nm) and the simulated resonant wavelength ($\lambda = 1326$ nm) is attributed to a slight deviation of structural parameter $s_0$ in device fabrication. Figure 3b shows the measured lasing intensity under different $P_{pump}$ (purple dots), which indicates a threshold pump power $P_{th} = 14.4$ μW. The linewidth-narrowing effect was also observed, which confirms the lasing operation. The lasing linewidth $\delta\lambda$ fitted from the measured spectra (orange squares) reduces from 2.66 to 1.23 nm as $P_{pump}$ increases to 160 μW. Figure 3c shows the measured spectrum (magenta open circles) and the corresponding Lorentzian fit (black solid line) at $P_{pump} = 14.2$ μW (just below the threshold), which indicate the linewidth $\delta\lambda = 1.46$ nm and the cavity $Q$ factor $\lambda/\delta\lambda = 920$. Compared with the simulated cavity $Q$ factor of 1590, the experimental $Q$ factor is lower, which can be attributed to unavoidable imperfections in device fabrication. As shown in Fig. 3d, increasing $P_{pump}$ leads to redshift in the peak wavelength of the Dirac-vortex laser with a linear coefficient of $d\lambda/dP_{pump} = 6.177$ nm/mW due to the thermal effect. In addition, we fabricated a series of Dirac-vortex lasers with varying $s_0$ and fixed $R_0 = a$ and $\delta_{max} = 35$ nm. The normalized lasing spectra in Fig. 3e indicate that the resonant wavelength of the topological Dirac-vortex lasers redshifts with a decreased $s_0$, which can be harnessed for tuning the lasing wavelength in a wide range of >70 nm.

The lasing properties of the Dirac-vortex cavity with different cavity sizes $R_0$ were further investigated. Figure 4a shows that the simulated resonant wavelength of the MBS is nearly independent of $R_0$, which is a unique feature of the topological MBS. The simulated $Q$ factor in Fig. 4a indicates that the Dirac-vortex cavity with a larger $R_0$ has a lower dissipation rate. Figure 4b shows the normalized experimental lasing spectra with different $R_0$ values, confirming that the lasing wavelength is always at ~1340 nm. The difference between the numerical (Fig. 4a) and experimental (Fig. 4b) lasing wavelength is attributed to the slight deviation of the structural parameter $s_0$ introduced in device fabrication. Note that two sidebands exist in the lasing spectrum of the Dirac-vortex cavity with $R_0/a_0 = 2$, as pointed by the black arrows in Fig. 4b. The FSR as determined from the separation in wavelength between the MBS and its nearest sideband is 31.8 nm. We also fabricated and measured Dirac-vortex cavities with $R_0/a_0 = 4$. By comparing the results with those of cavities with $R_0/a_0 = 2$, we found that the FSR is reduced by only 21.4% despite a nearly doubled modal volume (Extended Data Fig. 3). This clearly suggests that the FSR of the Dirac-vortex cavities possesses an unconventional scaling law with the modal volume $V$ that defies the $V^{-1}$ relation.

The far-field patterns and emission polarizations of the Dirac-vortex lasers with different cavity sizes $R_0$ were also investigated. Figure 4c and d show the measured and simulated $y$-polarized far-field patterns from the Dirac-vortex laser with $R_0/a_0 = 0.01$, suggesting that the $y$-polarized electric field component is zero. The white dashed circle indicates the numerical aperture (NA = 0.43) of the collection objective. Figure 4e and f show the measured and simulated $x$-polarized far-field patterns from the Dirac-vortex lasers with $R_0/a_0 = 0.01$, 0.5, 1, and 2. The experimental and numerical results agree well with each other, confirming that the Dirac-vortex lasers have pure linear polarization along the $x$ direction in their vertical emission in the far field. Figure 4e and f also suggest that the emission directionality is gradually enhanced with an increased cavity size $R_0$. A half width at half maximum of the emission beam of 8.6° (marked by the green dashed circles in Fig. 4e and f) can be obtained from the Dirac-vortex laser cavity with $R_0/a_0 = 2$.

In our experiment, the cavity size $R_0$ was constrained by two factors: the relatively weak in-plane optical confinement provided by the bandgap of the surrounding photonic crystal, and the limited size of the entire photonic crystal slab due to the suspended-membrane configuration. By adopting the same concept of this work, an electrically pumped Dirac-vortex laser with a much larger cavity size $R_0$ and without the need for a suspended photonic crystal slab can readily be realized, which is expected to have improved lasing performance, such as narrower lasing linewidth, smaller far-field divergence, and higher output power.

In conclusion, we experimentally demonstrated room-temperature continuous-wave Dirac-vortex topological lasers using InAs/InGaAs QD materials monolithically grown on a silicon substrate, obtaining single-mode linearly polarized vertical laser emission at a telecom wavelength. We confirmed that the lasing wavelength is topologically pinned to the Dirac point and that the FSR defies the universal $V^{-1}$ scaling law. These unique properties make our lasers fundamentally different from previous lasers with conventional cavity designs. These lasers are also the first topological lasers that are monolithically integrated on a silicon substrate and can operate under room-temperature continuous-wave conditions, representing an important step for integrating topological lasers on the silicon nanophotonic and microelectronic platform. Considering that photonic crystal surface-emitting lasers have recently been commercialized, our Dirac-vortex lasers will find various practical applications including transmitters in fiber-based or free-space communication networks, light sources in Lidar or face-recognition systems, and chemical or biomedical sensors. Meanwhile, as lasers inherently exhibit non-Hermiticity[28] and bosonic nonlinearity[29], our results enable further experimental exploration of new physics in the MBS that have no counterpart in the electronic domain. By reducing the density of the InAs/InGaAs QD such that each Dirac-vortex cavity contains only one QD, our devices provide an additional strategy for investigating the interplay between topology and quantum electrodynamics[30].

**Methods**

**Simulation.** Commercial software (COMSOL Multiphysics) was used to calculate the bulk states and their eigenfrequencies of the photonic crystal. Commercial software (Lumerical) was used to conduct three-dimensional finite-difference time-domain simulation of the Dirac-vortex cavity.

**Epitaxial growth of InAs/InGaAs QD on silicon.** The Dirac-vortex topological lasers were fabricated in InAs/InGaAs QD layers monolithically grown on an on-axis silicon (001) substrate. High-temperature annealing (900°C) with hydrogen atmosphere treatment was applied on a 300-mm-diameter silicon (001) substrate with a 0.15° misorientation in the [110] direction inside a metalorganic chemical vapor deposition system. A two-step 400-nm epitaxial GaAs film was grown to suppress the formation of antiphase boundaries. Then, the GaAs/silicon wafer was diced into 2-inch wafers for molecular beam epitaxy (MBE) growth. A 200-nm-thick GaAs buffer layer was deposited in the MBE chamber to achieve a smooth surface. Then, four sets of defect-filter layers were grown to reduce the density of threading dislocations owing to the large lattice mismatch between GaAs and silicon. Each defect-filter layer contains five repeats of $In_{0.18}Ga_{0.82}As$/GaAs strained-layer superlattice grown at 480°C and a 300-nm GaAs spacer layer grown at 590°C. The active layer was grown on a 1-μm-thick $Al_{0.6}Ga_{0.4}As$ sacrificial layer. The active material consists of four layers of InAs/$In_{0.15}Ga_{0.85}As$ dot-in-well, which are separated by 50-nm GaAs spacer layers and sandwiched between 40-nm $Al_{0.4}Ga_{0.6}As$ cladding layers. Each dot-in-well layer includes three monolayers of InAs deposited on a 2-nm $In_{0.15}Ga_{0.85}As$ quantum well and capped by a 6-nm $In_{0.15}Ga_{0.85}As$ layer.

**Device fabrication.** First, 120-nm-thick $SiO_2$ was deposited on the as-grown wafer using plasma-enhanced chemical vapor deposition. Second, the patterns of the photonic crystal were defined in resist ZEP520A by electron-beam lithography. Third, the patterns were transferred to the $SiO_2$ layer by using reactive-ion etching. Fourth, the residual resist was removed using $O_2$ plasma ashing. Fifth, the patterns of the photonic crystal were transferred from $SiO_2$ to the III–V materials through chlorine-based inductively coupled plasma reactive-ion etching. Then, the residual $SiO_2$ hard mask was removed by using diluted hydrofluoric acid. Finally, wet etching in 40% hydrofluoric acid solution was conducted to remove the 1-μm-thick $Al_{0.6}Ga_{0.4}As$ sacrificial layer to form an air cladding region underneath the photonic crystal slab.

**Measurement.** The fabricated devices were placed in a μ-PL measurement system with a surface-normal pump configuration at room temperature. A continuous-wave 632.8-nm He-Ne laser was used as the pump source. A 50× objective was used to focus the pump laser onto the center of the Dirac-vortex cavity, whose position was precisely controlled by piezoelectric nanopositioners. The emission from the devices was collected by the same objective. A long-pass filter was used to block the pump laser, and the emitted light was guided into a monochrometer with a thermoelectrically cooled InGaAs photodetector to characterize the emission spectra. A convex lens was used to project the emission from the device to a near-infrared camera, with the polarization state controlled by a linear polarizer, to characterize the far-field emission patterns.

**Data Availability** The data that support the findings of this study are available from the corresponding author upon reasonable request.

**Supplementary Information** is available in the online version of the paper.

**Acknowledgements** This work was supported by the Research Grants Council of Hong Kong (14209519), Shenzhen Key Laboratory Project (ZDSYS201603311644527), Optical Communication Core Chip Research Platform, UK Engineering and Physical Sciences Research Council (EP/P006973/1, EP/T01394X/1, EP/T028475/1), National Epitaxy Facility, European project H2020-ICT-PICTURE (780930), Royal Academy of Engineering (RF201617/16/28), French National Research Agency under the Investissements d'avenir ANR-10-IRT-05 and ANR-15-IDEX-02 and French RENATECH network. The devices were partially fabricated in the Core Research Facilities at Southern University of Science and Technology, whose engineers provided technical support.

**Author Contributions** J.M. developed the theoretical concept. J.M. conducted numerical simulation with assistance from X.X.. J.M. and T.Z. fabricated the devices with help from H.L. (Haochuan Li). T.Z. and J.M. conducted device characterization with help from Z.Z. (Zhan Zhang). T.Z. analyzed the experimental data. M.T., M.M., T.B., H.L. (Huiyun Liu), and S.C. performed epitaxial growth of the III–V-on-silicon wafer. J.M. and X.S. wrote the manuscript with input from all coauthors. Z.Z. (Zhaoyu Zhang), S.C., and X.S. supervised the project.

**Competing Interests** The authors declare no competing interests.

**Author Information** Reprints and permissions information is available at website. The authors declare no competing financial interests. Correspondence and requests for materials should be addressed to Z.Z. (zhangzy@cuhk.edu.cn), S.C. (siming.chen@ucl.ac.uk), and X.S. (xksun@cuhk.edu.hk).

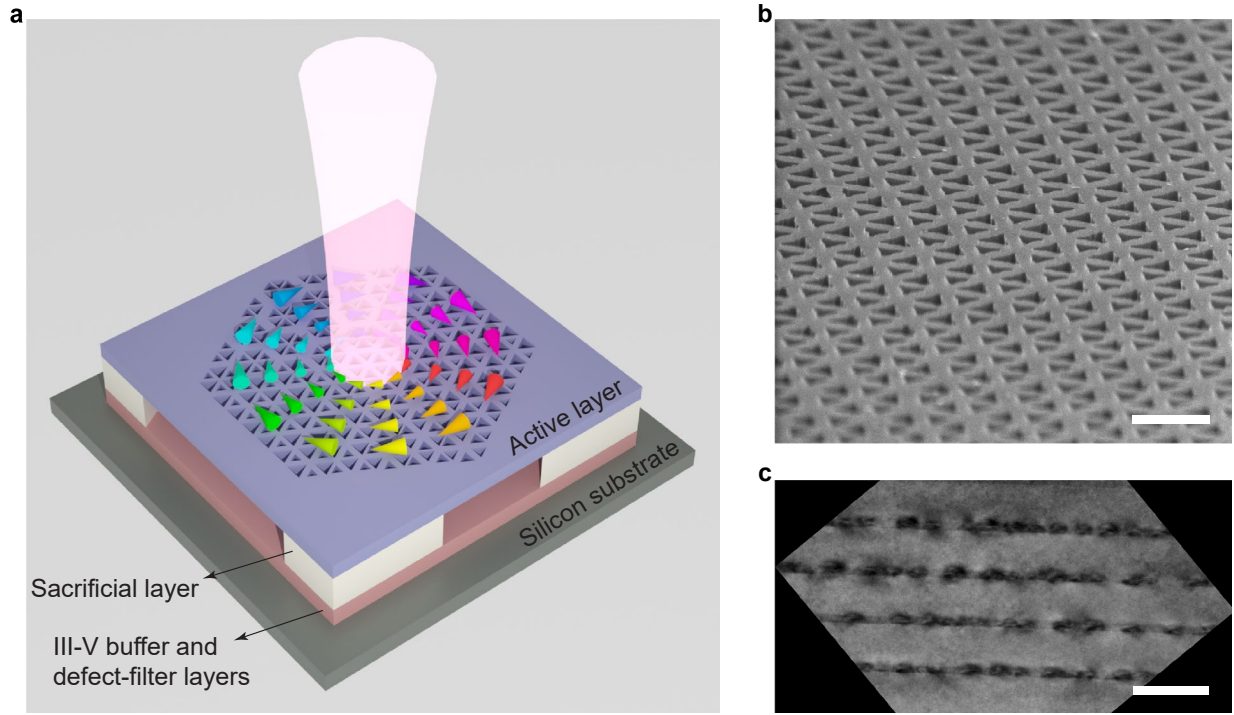

**Fig. 1 | Dirac-vortex topological lasers monolithically grown on silicon. a**, Conceptual illustration of a Dirac-vortex topological laser epitaxially grown on an on-axis silicon (001) substrate. The photonic crystal structure was defined in the active layer and suspended by partially removing the sacrificial layer. The III–V buffer and defect-filter layers were carefully optimized to minimize the effects of lattice mismatch between the III–V materials and silicon substrate. **b**, Tilted-view scanning electron microscope image of the fabricated topological Dirac-vortex photonic crystal cavity. Scale bar, 500 nm. **c**, Cross-sectional bright-field transmission electron microscope image of the active layer containing four-stack InAs/InGaAs QD layers. Scale bar, 100 nm.

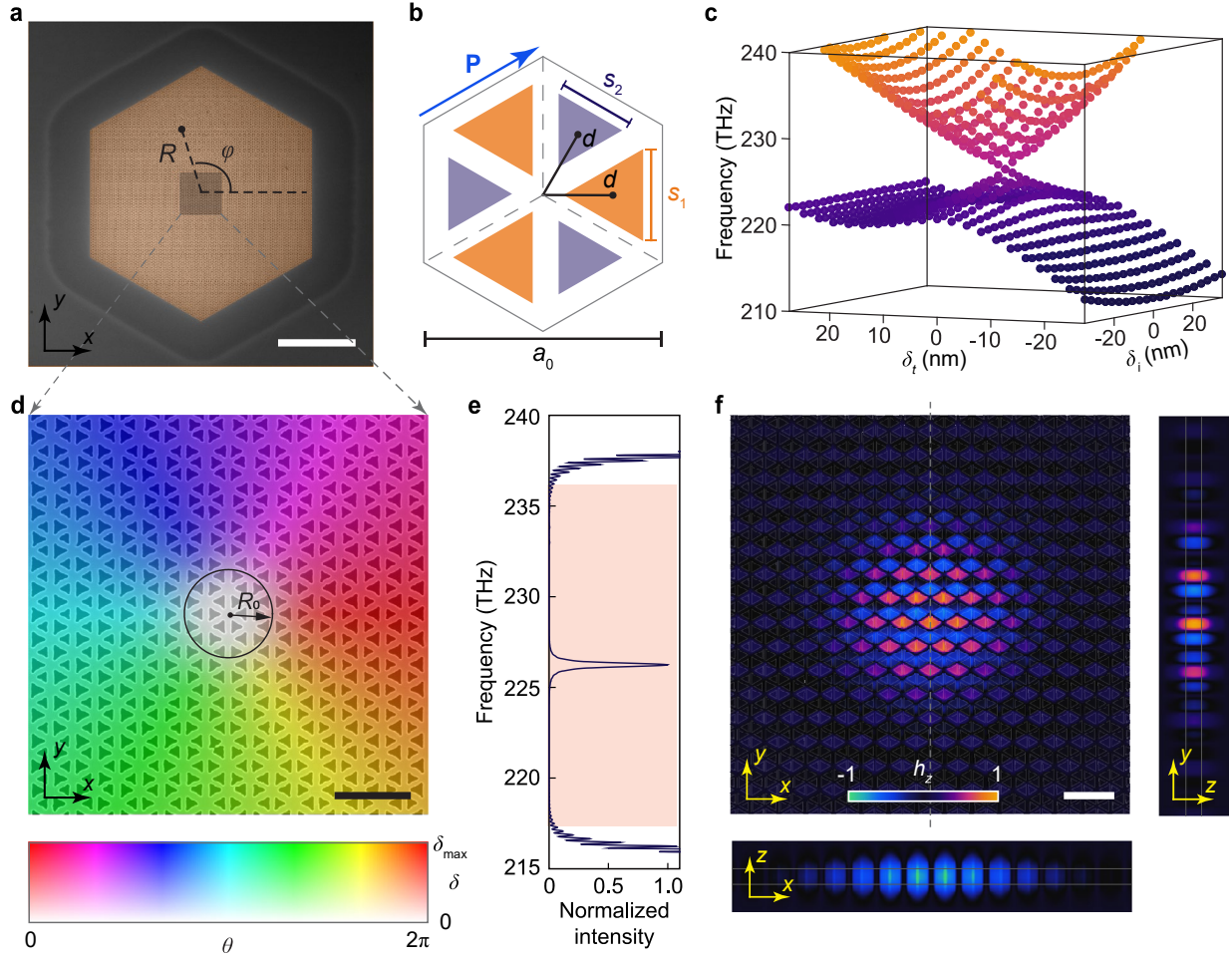

**Fig. 2 | Design and fabrication of the Dirac-vortex laser cavity. a**, Scanning electron microscope image of the fabricated Dirac-vortex topological photonic crystal laser. $R$ and $\varphi$ are the radial and angular coordinates, respectively. Scale bar, 10 μm. **b**, Illustration of the detailed structure in a unit cell. The lattice constant of the hexagonal photonic crystal is $a_0$ = 641 nm. Each unit cell contains six triangular holes that can be classified into two groups (colored in purple and orange). The side lengths of the two groups of triangular holes are governed by parameters $(s_1, s_2) = (s_0 + \delta_i, s_0 - \delta_i)$ with $s_0$ = 220 nm. The relative distance from the centers of these triangular holes to the center of the unit cell is $d = a_0/\sqrt{3} - \delta_t$. Here, $\delta_i$ breaks the $C_2$ inversion symmetry and $\delta_t$ breaks the $T_P$ translational symmetry along the vector **P** (marked by the blue arrow). **c**, Simulated eigenfrequencies of the bulk states at the Γ point of the first Brillouin zone with different values of $\delta_i$ and $\delta_t$. **d**, Scanning electron microscope image of the photonic crystal structure near the vortex center. It is color-coded by the spatially varying parameters $\delta_0(R) = \delta_{max} \cdot [\tanh(R/R_0)]^4$ and $\theta(\varphi) = \varphi$, where $R_0$ controls the size of the region with a near-zero value of $\delta_0(R)$. Scale bar, 1 μm. **e**, Simulated normalized intensity spectrum of the Dirac-vortex cavity. A topological Majorana bound state exists in the bulk bandgap (pink region). **f**, Simulated modal profiles ($h_z$ component) of the Majorana bound state. Scale bar, 1 μm.

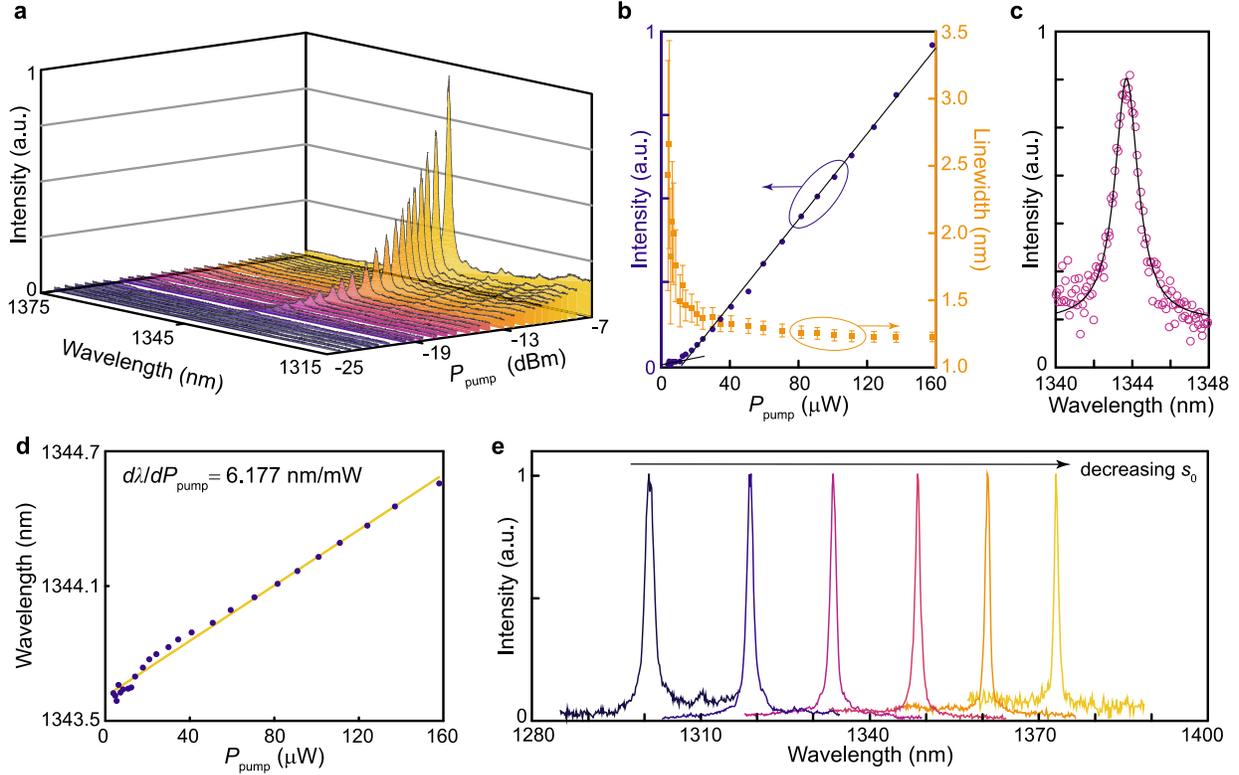

**Fig. 3 | Experimental characterization of the Dirac-vortex topological lasers. a**, Measured μ-PL spectra of the Dirac-vortex lasers with structural parameters $R_0 = 2a$ and $\delta_{max} = 35$ nm under different pump power $P_{pump}$. **b**, Measured μ-PL intensity (purple dots) and linewidth (orange squares) as a function of the pump power $P_{pump}$. The error bars represent the standard deviation in the linewidth fitting. The lasing threshold is $P_{th} = 14.4$ μW. **c**, μ-PL spectrum measured just below the lasing threshold ($P_{pump} = 14.2$ μW). The Lorentzian fit (black solid line) indicates a linewidth $\delta\lambda = 1.46$ nm. **d**, Measured lasing wavelength $\lambda$ (purple dots) as a function of the pump power $P_{pump}$, where a linear fit (orange line) suggests a linear coefficient of $d\lambda/dP_{pump} = 6.177$ nm/mW. **e**, Measured normalized lasing spectra from the Dirac-vortex lasers with varying $s_0$ and fixed $R_0 = a$ and $\delta_{max} = 35$ nm. The lasing wavelength can be tuned in a range wider than 70 nm.

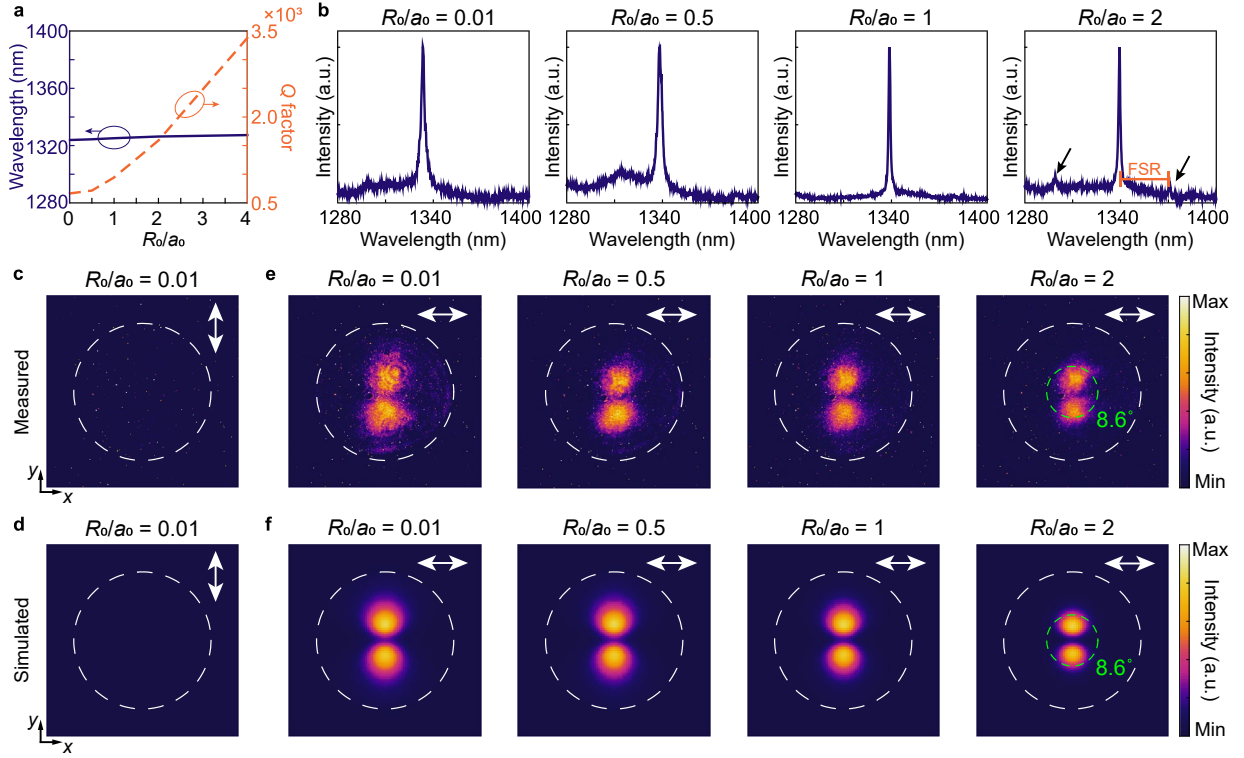

**Fig. 4 | Lasing characteristics of the Dirac-vortex topological lasers with different cavity sizes.**
**a**, Simulated cavity resonant wavelength (purple solid line) and $Q$ factor (orange dashed line) of the Dirac-vortex lasers with different cavity sizes $R_0$. **b**, Measured normalized lasing spectra from the Dirac-vortex lasers with $R_0/a_0 = 0.01$, 0.5, 1, and 2. Two sidebands (pointed by the black arrows) appear in the lasing spectrum of the device with $R_0/a_0 = 2$. The FSR is determined by the distance between the desired lasing peak and the undesired sidebands in the lasing spectrum. **c**, **d**, Measured (**c**) and simulated (**d**) $y$-polarized far-field emission patterns from the Dirac-vortex laser with $R_0/a_0 = 0.01$. **e**, **f**, Measured (**e**) and simulated (**f**) $x$-polarized far-field emission patterns from the Dirac-vortex lasers with $R_0/a_0 = 0.01$, 0.5, 1, and 2. Increasing $R_0$ leads to a decreased divergence angle. The Dirac-vortex laser with $R_0/a_0 = 2$ exhibits a half width at half maximum of the emission beam of 8.6°, as marked by the green dashed circles in **e** and **f**. The white double-headed arrows in **c**–**f** indicate the polarization direction for the measured far-field emission pattern. The dashed circles in **c**–**f** indicate the numerical aperture (NA = 0.43) of the collection objective.

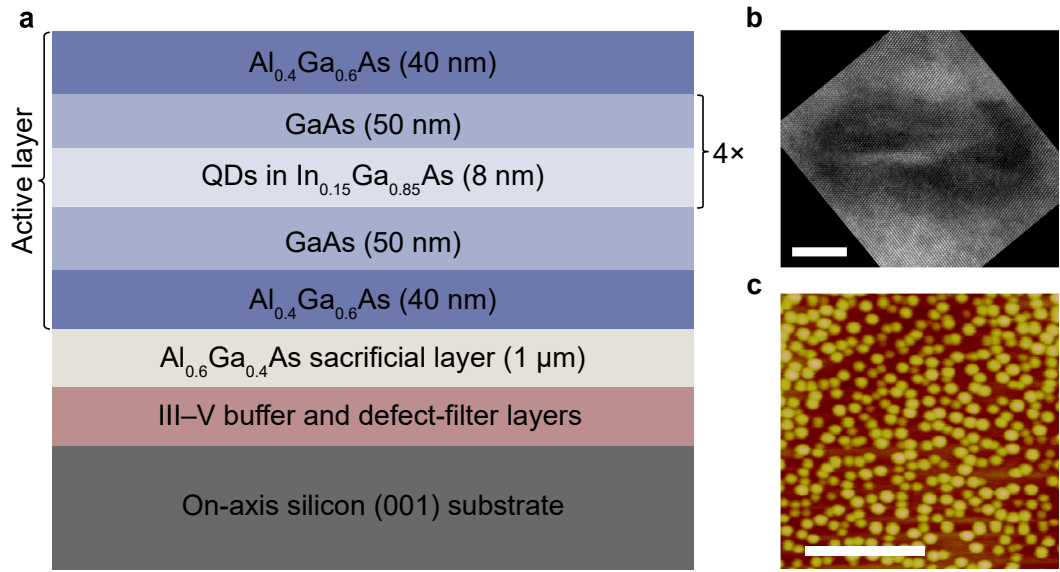

**Extended Data Fig. 1 | Epitaxial structure of the InAs/InGaAs QD active layer grown on silicon. a**, Illustration of epitaxial structure of the active layer, which consists of two symmetric 40-nm-thick $Al_{0.4}Ga_{0.6}As$ cladding layers and four layers of InAs/InGaAs dot-in-well structures separated by 50-nm GaAs spacer layers. **b**, Cross-sectional bright-field transmission electron microscope image of a single QD. Scale bar, 10 nm. **c**, Atomic force microscope image of uncapped InAs/InGaAs QDs. Scale bar, 400 nm.

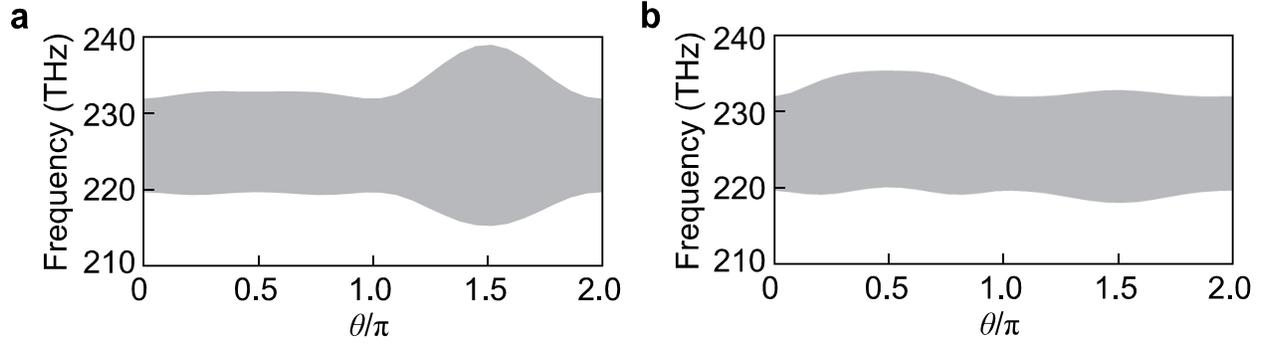

**Extended Data Fig. 2 | Dependence of the bulk bandgap on parameter $\theta$. a**, **b**, Simulated bulk bandgaps at the $\Gamma$ point of the first Brillouin zone with $\delta_0$ fixed at 35 nm and $\theta$ varying from 0 to $2\pi$. The geometric parameters $\delta_t$ and $\delta_i$ are determined by $(\delta_t, \delta_i) = \delta_0(\alpha \cdot \sin\theta, \cos\theta)$. The simulated bulk bandgap depends strongly on $\theta$ when $\alpha$ is 0.49 (**a**). In our experiment, we set $\alpha = 0.65$ (0.33) for $\delta_t > 0$ ($\delta_t < 0$) to obtain a weakly $\theta$-dependent bulk bandgap (**b**).

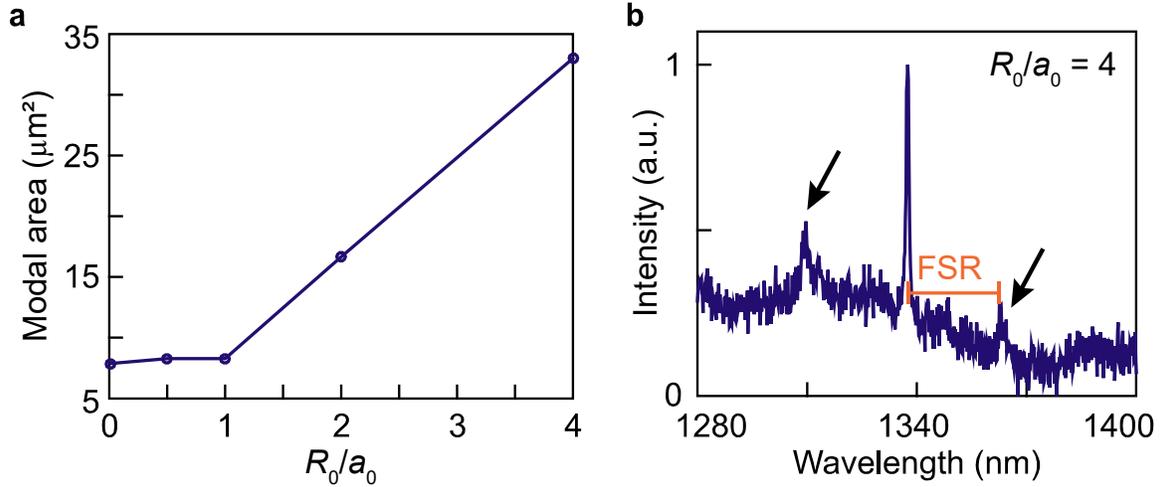

**Extended Data Fig. 3 | Unconventional FSR defying the universal inverse scaling law with the cavity size. a**, Simulated modal area of the Majorana bound state as a function of the cavity size $R_0$. **b**, Measured normalized lasing spectrum from a Dirac-vortex laser with $R_0/a_0 = 4$. The two sidebands (pointed by the black arrows) indicate an FSR of 25.1 nm, which is determined from the separation in wavelength between the MBS and its nearest sideband. Compared with the results from the topological laser with $R_0/a_0 = 2$, the FSR is reduced by only 21.4% while the modal area increases from 16.7 to 33.0 µm².